# Giant photon-drag-induced ultrafast photocurrent in diamond for nonlinear photonics

Xinyi Xue [a#], Wanyi Du [a#], Wei Tao [c,f #], Yuanyuan Huang *[a], Zhen Lei [a], Lipeng Zhu [b], Yuxiao Zou [f], Ying Liu [f], Gangqin Liu [c], Changzhi Gu [c], Yunliang Li *[c,d], Baogang Quan *[c,d,e], Xinlong Xu *[a]

a. Shaanxi Joint Lab of Graphene, State Key Laboratory of Photon-Technology in Western China Energy, International Collaborative Center on Photoelectric Technology and Nano Functional Materials, Institute of Photonics & Photon-Technology, Northwest University, Xi'an 710069, China

b. Xi'an University of Posts and Telecommunications, School of Electronic Engineering, Xi'an 710121, China

c. Beijing National Laboratory for Condensed Matter Physics, Institute of Physics, Chinese Academy of Sciences, Beijing 100190, China

d. School of Physical Sciences, University of Chinese Academy of Sciences, Beijing 100049, China

e. Songshan Lake Material Laboratory, Guangdong 523808, China

f. College of Chemistry, Beijing Normal University, Beijing 100875, China.

\# These authors contribute equally in this work.
\* Corresponding authors:
Yuanyuan Huang, E-mail: yyhuang@nwu.edu.cn
Yuliang Li, E-mail: yunliangli@iphy.ac.cn
Baogang Quan, E-mail: quanbaogang@iphy.ac.cn
Xinlong Xu, E-mail: xlxuphy@nwu.edu.cn

**Abstract**

Diamond is emerging as an attractive third-generation wide-bandgap semiconductor for future on-chip nonlinear photonics and quantum optics due to its unique thermal, optical, and mechanical properties. However, the light-driven current under below-bandgap excitation from the second-order nonlinear optical effect in diamond is still challenging. Herein, a giant second-order nonlinear photocurrent is observed in the chemical vapor deposition (CVD) diamond by utilizing terahertz (THz) emission spectroscopy. This ultrafast photocurrent originates from the photon drag effect (PDE), during which the momentum transfer from the incident photons to the charge carriers at the rich grain boundaries of the CVD diamond after the exclusive subgap π-π* transition upon femtosecond laser excitation. Especially, the interplay between circular and linear PDE to the THz generation has been clarified and distinguished under elliptically polarized light excitation. Furthermore, the picosecond ultrafast dynamics of these charge carriers are also verified by the infrared spectroscopy. Owing to the giant photon-drag-induced ultrafast photocurrent, the CVD diamond presents the highest THz emission efficiency compared with the reported carbon allotropes, which expands the new functionality of diamond nonlinear photonics into on-chip THz devices.

**Keywords:** CVD diamond, terahertz emission spectroscopy, nonlinear photocurrent, linear and circular photon drag effect, charge carrier

Carbon allotropes, owing to their unique combination of hybrid $sp^2$ and $sp^3$ orbitals, have created an attractive material family ranging from 1D carbon nanotubes, 2D graphene, to 3D graphite and diamond, which open a new era of carbon-based optoelectronics [1, 2]. Among these allotropes, diamond is viewed as the third-generation wide-bandgap semiconductor with extraordinary thermal, optical, and mechanical properties [3-5], which breeds so-called diamond photonics expected to make up and suppress silicon photonics in certain applications [6]. Especially, advances in diamond nonlinear photonics have provided a promising platform for frequency conversion [7], frequency microcombs at new wavelengths [8, 9], and pulse shaping [10]. The physical origin of these applications is based on the remarkable third-order nonlinearity of diamond including large optical Kerr effect [11], two-photon absorption [6], strong Raman scattering[12], and so forth. Such nonlinearity is generally treated as the lowest order nonlinear optical effect of diamond for decades[13], whereas the even-order nonlinearity of diamond still remains sparsely investigated owing to the inversion symmetry characteristics of diamond.

In order to activate the even-order nonlinearity of diamond, the color centers such as nitrogen-vacancy have been introduced in diamond to lift the inversion symmetry for the second harmonic generation [4, 13, 14]. However, there still remain questions from two aspects. On the one hand, it is challenging to monitor and control the large-scale diamond due to the atomic-scale color center sizes from the material-growth processing point of view[15]. On the other hand, light-driven photocurrent in diamond stemming from second-order nonlinear effects becomes increasingly important for optoelectronic devices[13, 16], but the fundamental physics of the nonlinear photocurrent in diamond has been veiled so far. Intuitively, diamond is electrical insulating with a large bandgap (5.5 eV) [17] and offers a wide transparency window [18] due to the special hybrid orbital $sp^3$, which makes it impossible for diamond to generate photocurrent under Vis-infrared light excitation. Intriguingly, chemical vapor deposition (CVD) diamond enables large-scale diamond synthesis as well as manipulates free carriers and inversion symmetry of diamond by increasing the $sp^2$ bonding fraction [19]. Benefiting from these characteristics, CVD diamond not only presents excellent electrical conductivity (~100 S/cm) [20] and even superconductivity [21, 22], but also enriches the nonlinear photonics spanning from the superior nonlinear refractive index (n=1.3×10$^{-19}$) [23] to up-converted photoluminescence[24] and two-photon absorption [25]. Despite of these progresses, the second-order nonlinear photocurrent in CVD diamond is still in challenge. Especially how to generate, control, and detect the second-order nonlinear optical response of the carrier dynamics in diamond needs to be explored, which is crucial to extend the diamond nonlinear photonics and on-chip optoelectronic devices towards new wavelengths such as telecom and terahertz (THz) regime.

In this work, we report a giant second-order nonlinear photocurrent in a millimeter-scale CVD diamond film via the THz emission spectroscopy. The linearly and elliptically polarized femtosecond (fs) pulse excitation would lead to linear momentum and spin angular momentum transfer from the incident photons to the charge carriers due to the subgap π-π* transition, which results in linear photon drag effect (LPDE) and circular photon drag effect (CPDE), respectively. Intriguingly, the CVD diamond exhibits the strongest THz emission efficiency (~1.6×10$^9$ V/J) among the reported carbon allotropes based on the high laser threshold and primarily the presence of a large number of charge carriers at the rich grain boundaries. Furthermore, the ultrafast dynamics of these charge carriers are verified by the infrared spectroscopy.

## Results and Discussion

### Linear photon drag effect in CVD diamond

The diamond film is prepared by a hot-filament chemical vapor deposition (HFCVD) method, which is composed of diamond grains in micrometer scales (see Supplementary Note 1). The schematic of the THz emission process is depicted in Fig. 1a. A pump light with 800 nm central wavelength is used to illuminate the CVD diamond under an oblique incidence ($\theta$=40º), and the emitted horizontally- ($E_{THz-p}$) and vertically-polarized component ($E_{THz-s}$) of the THz electric field are obtained via a pair of wire-grid polarizers (see details in the Method section). Note that the conducting surface of the CVD diamond is ~50 nm in thickness, which contains rich grain boundaries on the insulating bulk with ~48 μm thickness (Fig. 1a). Using the THz emission spectroscopy (see Supplementary Note 2), the THz time-domain waveforms of the CVD diamond and a monocrystalline diamond (size ~5×5×0.5 mm$^3$) are compared as shown in Fig. 1(b). It is noteworthy that a significant THz wave is emitted from the CVD diamond, while no obvious THz wave is generated from the monocrystalline diamond. This phenomenon results from the good insulation characteristic of the monocrystalline diamond, which has no charge carriers as all valence electrons participate in the covalent bond formation. On the one hand, the bandgap of monocrystalline diamond [26] is much higher than that of the photon energy of pump light (~1.55 eV). As such, no THz-signal in monocrystalline diamond suggests there is no multi-photon absorption process to induce photocarriers[27, 28]. On the other hand, different from the typical THz emitter such as ZnTe (bandgap ~2.28 eV)[29], the optical rectification that can be induced under below-bandgap excitation in the monocrystalline diamond is also ruled out. For the CVD diamond, its absolute value of the generated THz electric field ($|E_{THz}|$) is approximately 1702 V/m (see details in Supplementary Note 3), which is 12 and 243 times larger than that of vertical grown graphene (VGG, ~142.46 V/m)[30] and multilayer graphene (MG, ~7 V/m)[31]. Besides, the bandwidth of THz frequency-domain spectrum of CVD diamond (~2.5 THz) is slightly wider than that of MG[31] and VGG[30] as shown in Fig. 1(c). We have summarized the THz emission efficiency ($\eta$) of different materials to evaluate the practical potential of CVD diamond for THz emitter as shown in Fig. 1(d). Notably, the CVD diamond presents a giant THz emission efficiency (~$1.6\times10^9$ V/J) per unit thickness, which is superior to the reported carbon allotropes including 1D multi-wall carbon nanotubes (MWNT)[32], 2D VGG and MG[30, 31], and 3D graphite[33], and the efficiency is even two orders of magnitude higher than that of the standard THz emitter such as ZnTe (~$6.9\times10^7$ V/J)[34]. Furthermore, as shown in Fig.1 (e), the peak-to-valley values of both $E_{THz-p}$ and $E_{THz-s}$ are independent on the azimuthal angle of the CVD diamond due to the isotropy and randomness of grain orientations on the CVD diamond surface. This azimuthal angle-independent feature is different from the traditional THz emitters such as ZnTe and GaAs[35, 36], which induce azimuthal angle-sensitive THz emission due to their nonlinear susceptibility determined by the monocrystalline symmetry. In this regard, the uniformity of the polycrystalline CVD diamond is beneficial for uniform THz emission without azimuthal angle dependence. As a result, the CVD diamond exhibits great potential as high-efficiency on-chip THz emitters among the third-generation semiconductors.

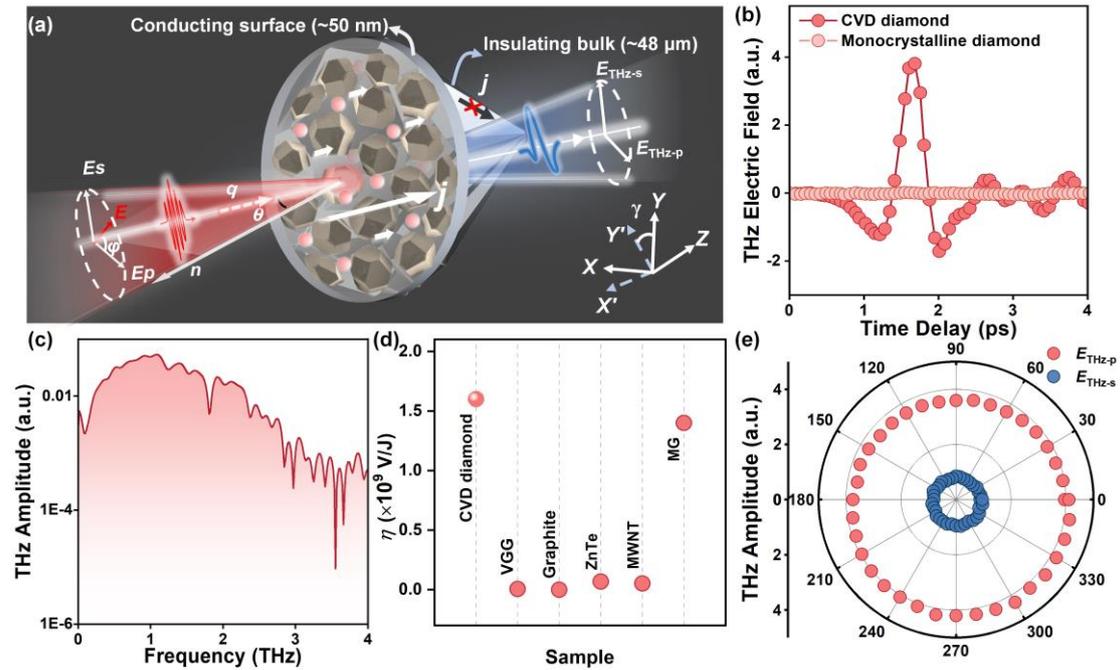

**Fig. 1 Remarkable THz emission properties of the CVD diamond.** (a) The simplified diagram of the THz emission of the CVD diamond with prescribed

laboratory (*XYZ*) and crystalline (*X'Y'Z'*) coordinates. The incident angle ($\theta$) is described as the angle between the incident light and the normal (***n***) direction of the CVD diamond. (b) THz time-domain electric fields of the CVD diamond (red) and monocrystalline diamond (pink) under 800 nm laser excitation; (c) The THz frequency-domain spectrum of the CVD diamond. (d) Comparison of the THz emission efficiency ($\eta$) between different materials. (e) Azimuthal angle dependence of the generated $E_{\text{THz-p}}$ (red) and $E_{\text{THz-s}}$ (blue) of the CVD diamond.

In order to elucidate the THz emission mechanism of the CVD diamond, we investigate the THz peak-to-valley amplitude with respect to the pump fluence under the oblique incidence. Two different evolutions of the pump-dependent THz amplitude are shown in Fig. 2(a): (I) Both $E_{\text{THz-p}}$ and $E_{\text{THz-s}}$ of the CVD diamond increase linearly as the pump fluence increasing from 0 mJ/cm² to 2.12 mJ/cm² (see the red region in Fig. 2(a)). This linear dependence of the THz amplitude confirms the existence of a second-order nonlinear optical effect as $E_{\text{THz}} \propto \chi^{(2)} |E(\omega)|^2 \propto I_{\text{pump}}$ [37]. (II) Specially, above a high pump fluence (>2.12 mJ/cm², see the blue region in Fig. 2(a)), both $E_{\text{THz-p}}$ and $E_{\text{THz-s}}$ present the amplitude saturation with the pump fluence, which results from the limited amount of charge carriers at the conducting surface in the nanometer-scale thickness. This saturated pump fluence dependence is different from the reported MoSe$_2$[38] and MoS$_2$[33], which originate from the electrostatic screening and the surface damage, respectively. It is noteworthy that even the pump fluence is up to 4.2 mJ/cm², the CVD diamond consistently emits THz emission without any surface damage, which could be ascribed to the high thermal conductivity in the CVD diamond[39]. This result indicates that the CVD diamond has a high-power damage threshold, which is promising for integrated optical components in THz technology, nonlinear optics, and high energy physics.

In general, both linear and nonlinear optical processes can induce ultrafast photocurrent for THz emission. The surface photocurrent induced by linear optical effects includes the surface depletion field effect and the photo-Dember effect, which has been excluded in our experiments as discussed in Supplementary Note 4. As for the nonlinear optical processes, the photocurrent can be introduced via the photon drag effect and photogalvanic effect. For the photogalvanic effect, when electrons are excited from the valence band to the conduction band, the charge centers move along the polar direction, leading to the generation of photocurrents (so-called shift current) [40]. This is a second-order nonlinear process that requires inversion symmetry broken[41], which can also be ruled out since the CVD diamond exhibits surface isotropy according to its azimuthal angle dependence of THz amplitude in Fig. 1(e). As for the photon drag effect, its physical origin is the momentum transfer from the incident photons to the carriers, which is directly related to the incident wave-vector and could occur on any centrosymmetric materials in principle[42].

With the aim of verifying the dominant role of the photon drag current, we compare the THz time-domain signals of the $E_{\text{THz-p}}$ and $E_{\text{THz-s}}$ illuminated on the front and back sides, which is equivalent to the incident light wave-vector reversal with respect to the CVD diamond. Obviously, the polarity of THz time-domain waveforms of both $E_{\text{THz-p}}$ and $E_{\text{THz-s}}$ are reversed when the CVD diamond is illuminated from the opposite sides as shown in Figs. 2(b-c). This wave-vector dependent THz emission is a typical characteristic of the photon drag effect, confirming a predominant mechanism for the THz emission. Besides, this process is also verified by the incident angle dependence of the $E_{\text{THz-p}}$ and $E_{\text{THz-s}}$ as shown in Fig. 2(d). The $E_{\text{THz-p}}$ is an odd function of the incident angle, while the $E_{\text{THz-s}}$ has a slight change at different incident angles under p-polarized light excitation (with the polarization angle *φ=0*). Moreover, under the normal incidence (*θ=0º*), the THz emission of both $E_{\text{THz-p}}$ and $E_{\text{THz-s}}$ are almost 0, which could be ascribed to two aspects. One is the symmetric distributions of transient electrons and holes in the momentum space. As a result, their contributions for photon drag current are canceled out along opposite momentum directions, whereas a net current is generated at oblique incidence due to the asymmetry of carrier distributions [31]. The other is the disappearance of in-plane wave vector projection[42] under normal incidence as shown in Fig. S4b (see Supplementary Note 5). Therefore, only out-of-plane wave vector contributes to the THz emission, which cannot be detected along the propagating direction (Z-axis) (Fig. S5).

Space symmetry of the CVD diamond is directly related to the nonlinear response of the photocurrent since the nonlinear conductivity tensor is determined by the space symmetry. Different from the monocrystalline diamond ($m\bar{3}m$), the CVD diamond surface exhibits random atomic arrangement in a macroscopic view (Fig. 1(e)). Therefore, the CVD diamond exhibits ∞m point group, which is considered for analyzing the nonlinear photocurrent in CVD diamond [43]. In the experiment, the photocurrent induced THz emission presents evident dependence on the polarization angle of the pump laser. As shown in Figs. 2(e-f), both $E_{\text{THz-p}}$ and $E_{\text{THz-s}}$ exhibit $2\varphi$-rotational symmetry when $\varphi$ increases from 0° to 360°. During this process, the $E_{\text{THz-p}}$ exhibits the same THz time-domain waveform polarity accompanied with an upward offset of THz amplitude (as indicated by the red arrow in Fig. 2(e)), while the $E_{\text{THz-s}}$ presents polarity modulation under different incident polarization angles. The distinguishing response between the $E_{\text{THz-p}}$ and $E_{\text{THz-s}}$ can be understood based on the phenomenological expression of the photocurrent density induced by LPDE ($J_\lambda^{LPDE}$) as follows[44]:

$$j_\lambda^{\text{LPDE}} = T_{\lambda\eta\mu\nu} q_\eta E_\mu E_\nu^* \tag{1}$$

where $T_{\lambda\eta\mu\nu}$ is the fourth-order nonlinear tensor; $q_\eta$ is the wave vector; $E$ is the incident electric field. The dependences of the THz components on the polarization angle are written as follows [30, 45] (see details in Supplementary Note 6):

$$E_{\text{THz-p}} = A_{pol} \cos 2\varphi + B_{pol} \tag{2}$$

$$E_{\text{THz-s}} = C_{pol} \sin 2\varphi \tag{3}$$

where the fitting constant $A_{pol}$ is related to $T_{x\eta\mu\nu}$ and $T_{z\eta\mu\nu}$, and $C_{pol}$ is only related to $T_{y\eta\mu\nu}$. Note that the constant term $B_{pol}$ related to $T_{x\eta\mu\nu}$ and $T_{z\eta\mu\nu}$ dominates the upward offset (Fig. 2(e). The experimental results can be well fitted with Eqs. (2-3) as shown in Figs. 2(e-f). From these fitting constants, $|A_{pol}|$ is comparable to $|C_{pol}|$, resulting in similar THz amplitude modulation for $E_{\text{THz-p}}$ and $E_{\text{THz-s}}$. Furthermore, combined with the dipole radiation theory and LPDE model, the dependence of THz emission on the incident angle under p-polarized excitation (see Supplementary Note 7) is calculated as: $E_{THz-p}=A_{inc}\cos 2\theta + B_{inc}\sin 2\theta + C_{inc}$. The fitting constants $A_{inc}$ and $C_{inc}$ are both related to $T_{xxxx}$, $T_{xxzz}$, $T_{zzxx}$, and $T_{zzzz}$, while the fitting constant $B_{inc}$ is related to $T_{xzzx}$ and $T_{zxxz}$. The incident angle dependence of the $E_{THz-p}$ is also in a good agreement with the theoretical results as shown in Fig. 2(d). Based on the fitting constants in Eqs. S14 in Supplementary Note 7, we can estimate that $|T_{xxxx}|$ is much higher than $|T_{zzxx}|$. Combined the THz amplitude dependences on the incident angle and polarization angle of the pump light, the LPDE is confirmed to be the dominant mechanism for the THz emission of CVD diamond under the linearly polarized excitation.

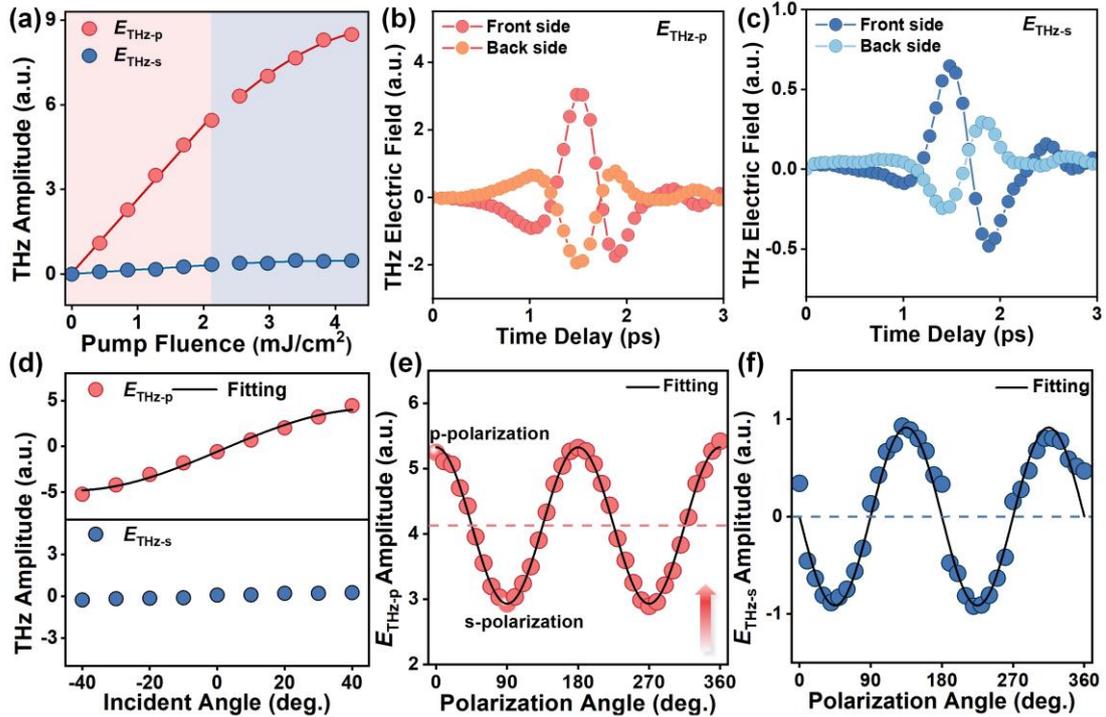

**Fig. 2 Linear photon drag current induced THz emission of the CVD diamond under linearly polarized light excitation.** (a) THz amplitude of the CVD diamond with respect to the pump fluence (red: $E_{THz-p}$; blue: $E_{THz-s}$). The time-domain waveforms of (b) $E_{THz-p}$ and (c) $E_{THz-s}$ excited from the front side and back side. The polarization angle $\varphi$ is fixed at 45° to simultaneously obtain $E_{THz-p}$ and $E_{THz-s}$. (d) The incident angle dependence of the $E_{THz-p}$ and $E_{THz-s}$ under p-polarized light excitation. The incident polarization angle dependence of the (e) $E_{THz-p}$ and (f) $E_{THz-s}$. The points

and lines are the experimental data and fitting results, respectively.

**Circular photon drag effect in CVD diamond**

Different from the linearly polarized pump light, the elliptically and circularly polarized pump light has both linear momentum and spin angular momentum, which could intriguingly result in the spin angular momentum transfer from the incident photons to the charge carriers in CVD diamond [46]. To investigate this so-called CPDE, a rotatable quarter-wave plate (QWP, $\alpha$) is added to change the polarization states of the pump light. Figs. 3(a-b) present the THz time-domain waveforms generated from the CVD diamond under left circularly polarized (LCP, $\alpha=45°$), right circularly polarized (RCP, $\alpha=135°$), and linearly polarized (LP, $\alpha=0°$) light excitations. Note that the generated $E_{\text{THz-p}}$ has the same polarity under LCP, RCP, and LP excitations. Besides, the THz amplitude under LP excitation is 1.2 times larger than that under LCP and RCP excitations as shown in Fig. 3a. However, the generated $E_{\text{THz-s}}$ has a similar amplitude and the same polarity under RCP and LP excitations, while the THz amplitude is almost 0 under the LCP excitation as shown in Fig. 3b.

To compare the THz amplitude evolution, the THz peak-to-valley amplitude with respect to the QWP angle are shown in Figs. 3(c-d), where both $E_{\text{THz-p}}$ and $E_{\text{THz-s}}$ present $2\alpha$-rotational symmetry as changing the QWP angle from 0º to 180º. These results can be understood based on the nonlinear photocurrent generation under elliptically polarized light excitation, at which the LPDE and CPDE can simultaneously contribute to the photocurrent density ($J_\lambda^{\text{PDE}}$) as[46]:

$$j_\lambda^{\text{PDE}} = T_{\lambda\eta\mu\nu}q_\eta E_\mu E_\nu^* + \tilde{T}_{\lambda\eta\nu}q_\eta P_{circ}\hat{e}_\nu I \qquad (4)$$

where the fourth-order nonlinear tensor $T_{\lambda\eta\mu\nu}$ determines the LPDE, while the third-order pseudo-tensor $\tilde{T}_{\lambda\eta\nu}$ determines the CPDE. The parameters $\hat{e}_\nu$ and $I$ are the unit vector and intensity of the incident light, respectively. The $P_{circ} = \sin 2\alpha$ represents the circular polarization degree. Combined with the Jones matrix and the dipole radiation theory, the THz electric field can be expressed as (see details in

Supplementary Note 8):

$$E_{\text{THz-p}} = L_x \cos 4\alpha + C_x \sin 2\alpha + D_x \tag{5}$$

$$E_{\text{THz-s}} = L_y \sin 4\alpha + C_y \sin 2\alpha \tag{6}$$

where the fitting constants $L_x$ and $L_y$ are related to $T_{\lambda\eta\mu\nu}$, while $C_x$ and $C_y$ are only related to $\tilde{T}_{x\eta\nu}$. The constant term $D_x$ represents the QWP-angle-independent background (BG) photocurrent, which is related to $T_{x\eta\mu\nu}$ and $T_{z\eta\mu\nu}$. Based on the fitting constants $A_k$, $B_k$, $L_k$, $C_k$, and $D_k$ ($k=$ *pol*, *inc*, *x*, *y*) by Eqs. (2-6), the fourth-order nonlinear tensor $T_{yxxy}$ is calculated as 2.8 times higher than $T_{yzzy}$. Furthermore, the contribution ratio of LPDE and CPDE in $E_{\text{THz-p}}$ is calculated as 93% and 7%, while the ratio in $E_{\text{THz-s}}$ is calculated as 84% and 16% (see details in Supplementary Note 8). The difference of the CPDE contribution ratio is related to the nonlinear coefficients as the $E_{\text{THz-s}}$ is determined by $\tilde{T}_{yzx}$ and $\tilde{T}_{yxz}$, while $E_{\text{THz-p}}$ is determined by only $\tilde{T}_{yzx}$ (see Eq. (S20) in Supplementary Note 8). Besides, based on the theoretical fitting by Eqs. (5-6), the third-order pseudo-tensor $\tilde{T}_{yzx}$ is 4 times higher than $\tilde{T}_{yxz}$. From the physical mechanism view, the coexistence of both LPDE and CPDE also indicates the effective transfer of linear momentum and spin angular momentum to the charge carriers at grain boundaries on the CVD diamond surface.

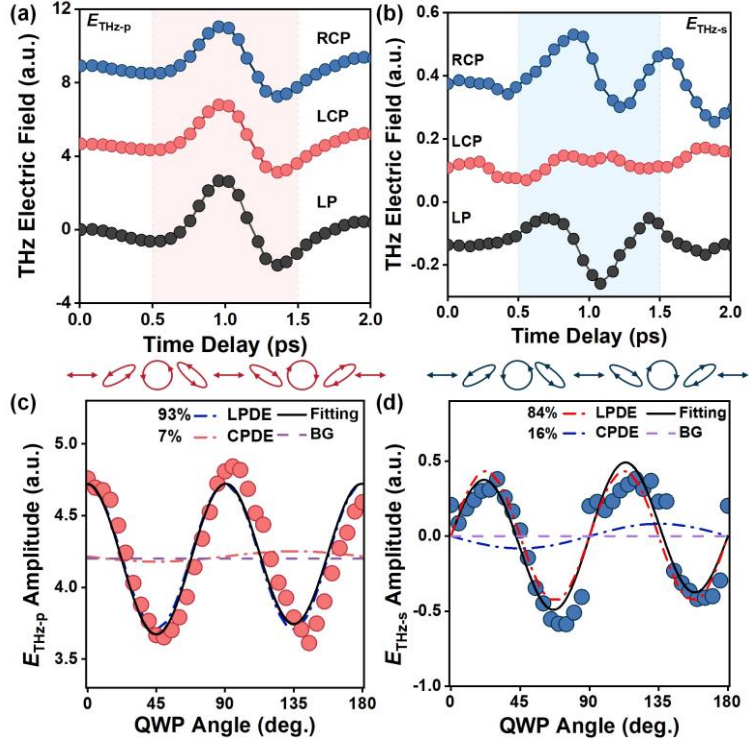

**Fig. 3 THz emission characteristic of the CVD diamond under elliptically polarized light excitation.** The time-domain (a) $E_{THz\text{-}p}$ and (b) $E_{THz\text{-}s}$ generated from the CVD diamond under LP (black), LCP (red), and RCP (blue) excitations. The peak-to-valley values of (c) $E_{THz\text{-}p}$ and (d) $E_{THz\text{-}s}$ with respect to the QWP angle. The polarization states of the excitation laser are depicted at the top of the diagram. The points, solid lines, and dashed lines are the experimental values, fitting results, and the contributions of LPDE, CPDE, and BG photocurrents, respectively.

**Carrier dynamics in CVD diamond**

As a proof of the mechanism for PDE, Fig. 4(a) depicts the static infrared transmission spectrum of the CVD diamond. The peaks located at 2734 cm$^{-1}$ and 2857 cm$^{-1}$ correspond to the stretching vibration frequencies of C(111)-H[47]. Besides, the characteristic peak at 2911 cm$^{-1}$ corresponds to the stretching vibration frequency of C(100)-H[48]. Note that the infrared transmission peak located at 2798 cm$^{-1}$ originates from the unintentional boron atom doping during the growth process[49]. Therefore, the C-H bonds on the CVD diamond surface could be related to the unpaired electrons in the sp$^2$ carbon and bonded H atoms in the ambient environment. As such, there are a large number of charge carriers and C-H dangling bonds exist at rich grain boundaries of the CVD diamond, which is consistent with the aforementioned discussion of LPDE and CPDE. According to the Drude-Lorentz model, the infrared transmission spectrum can be well fitted by Eq. S24 as shown in Fig. 4(a) (see Supplementary Note 9). Based on the dielectric function, the refractive index ($n$) and extinction coefficient ($\kappa$) have been extracted from the transmission spectrum (Fig. 4(a)). As shown in Figs. 4(b-c), the value of $n$ varies from 0.8 to 2.5, while $\kappa$ is in the order of $10^{-2}$, indicating that the CVD diamond still exhibits excellent infrared transparency[50]. Moreover, according to Eq. S25, the charge carrier concentration on the CVD diamond is calculated as $2.31 \times 10^{18}$ cm$^{-3}$, which is higher than that of the p-doped graphene ($\sim 3 \times 10^{12}$ cm$^{-3}$)[51]. Besides, the carrier mobility $\mu$ of the CVD diamond is calculated as 1494 cm$^2$V$^{-1}$·s$^{-1}$, which is in the same scale to the electron ($\sim$3800 cm$^2$V$^{-1}$·s$^{-1}$) and hole ($\sim$4500 cm$^2$V$^{-1}$·s$^{-1}$) mobility of the plasma-deposited diamond[52] (see details in Supplementary Note 10).

The dynamic process of charge carriers on the CVD diamond surface is investigated via the time-resolved transient mid-IR spectroscopy with pump-probe scheme[53]. The pulse centered at 800 nm with 100 fs pulse duration is used as a pump beam, and the mid-IR pulses centered at 3676 nm and 3534 nm are used as the probe beams to investigate the first and second excited states of the C-H bond [54], respectively. The dynamics of transmission transients at 3534 nm under different pump fluences are displayed in Fig. 4(d), and the results for 3676 nm are listed in Fig. S7 in Supplementary Note 11 as references. It displays that the transmission increases immediately after photoexcitation following a slower decay. The peak value of transmission changes as a function of the pump fluence as plotted in Fig. 4(e), which it demonstrates the linear absorption of carriers and rules out the nonlinear effects of two-photon and multi-photon pump absorption in our measurements[55]. Since the bandgap of the CVD diamond is much larger than the photon energy of the pump light[56], this distinguishes from the interband transition in other carbon allotropes (e.g. graphene and carbon nanotubes). Therefore, the optical absorption here should be connected with the sub-bandgap π-π* transition that is dominated in polycrystalline CVD diamond (Fig. 4(g-I)) and easily occurs on the grain boundary regions [57].

For the transmission decay process, the corresponding decay curve can be fitted with a double exponential function, and the time constants of the fast ($\tau_1$) and slow ($\tau_2$) processes varying with the pump fluence are listed in Fig. 4(f). The values of fast and slow time decay constants ($\tau_1$~0.3 ps and $\tau_2$~2 ps) are similar to the previous measurements on ultrafast carrier dynamics of different graphene materials [58-62]. It can be found that the time decay constants are almost independent on the pump fluence, which implies the dominant contributions to $\tau_1$ and $\tau_2$ should not be decided by carrier densities and are unlikely related to carrier-carrier scattering [55, 63]. Therefore, the fast and slow dynamics of transmission decay can be dominated by carrier-phonon scattering (Fig. 4(g-II)) and phonon-assisted charge carrier recombination (Fig. 4(g-III)), respectively [64]. Accordingly, the photocurrent would be generated with the three consecutive processes as depicted in Fig. 4(g), and this dynamic process is also consistent with the generation of photon drag current and THz emission from the CVD diamond on the picosecond-scale (Fig. 1(b)). Moreover, the photon drag effect requires the participation of the lattice (i.e. carrier-phonon interaction) for momentum transfer from incident photons to charge carriers [65], which could be enhanced due to the strong carrier-phonon scattering in the CVD diamond. As such, we have shed light on the physical relevance between the charge carriers and the ultrafast photocurrent as well as the subsequent THz emission of the CVD diamond.

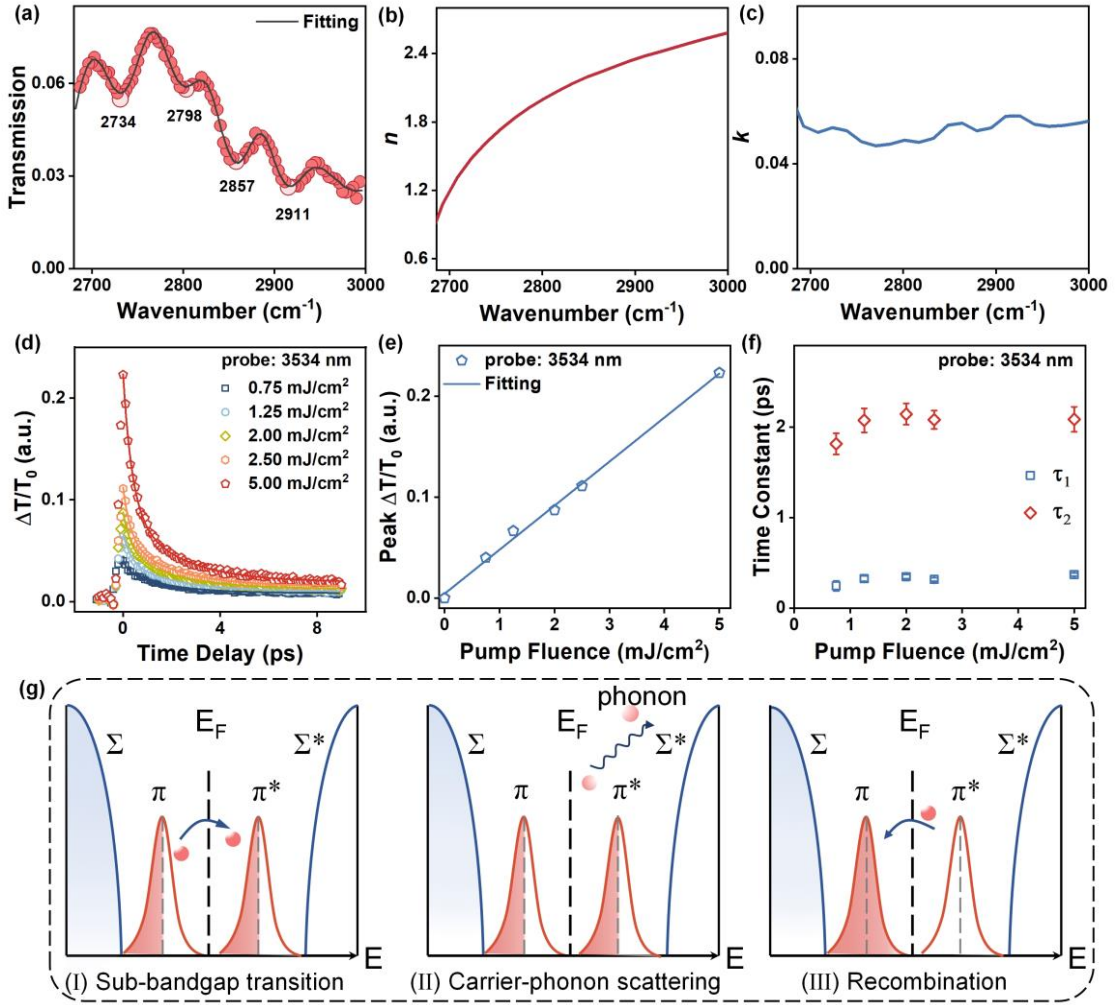

**Fig. 4 The static and dynamic infrared transmission spectra of the CVD diamond.** (a) Static infrared transmission spectra of the CVD diamond. (b-c) The refractive index (*n*) and extinction coefficient (*κ*) of the CVD diamond. (d) Transient infrared transmission spectroscopy of the CVD diamond probed under 3534 nm infrared light. The dependence of (e) the transient transmission peak and (f) the time constants of the fast ($\tau_1$) and slow ($\tau_2$) processes on the pump fluence. The points represent the experimental results and the solid lines represent the fitting results. (g) The carrier dynamics of the CVD diamond: (I) sub-bandgap π-π* transition, (II) carrier-phonon scattering, and (III) carrier recombination.

**Conclusion**

In summary, the second-order nonlinear response of ultrafast photocurrent is activated in the CVD diamond film observed via THz emission spectroscopy. The primary THz generation mechanisms are ascribed to the LPDE and CPDE, which originate from the linear momentum and spin angular momentum transfer from the incident photons to the charge carriers at rich grain boundaries on the diamond surface, respectively. The charge carrier concentration ($2.31 \times 10^{18}$ cm$^{-3}$) and picosecond-scale carrier dynamics of the diamond are obtained by the infrared spectroscopy. Notably, among the carbon allotropes, diamond exhibits a giant THz emission efficiency ~$1.6 \times 10^9$ V/J per unit cell thickness, which fills the new piece of the jigsaw puzzles in carbon family for THz emission. Furthermore, the interplay between CPDE and LPDE has been clarified and distinguished based on the THz amplitude dependence on the incident polarization angle. This work not only realizes generation and manipulation of the ultrafast nonlinear photocurrent in diamond, but also propels the diamond nonlinear photonics in micro-chip THz emitters.

**Methods**

**Sample Preparation**

The polycrystalline diamond film was grown on a 4-inch n-doped Si wafer using a HFCVD method. First, a Si wafer was sonicated in a nanodiamond suspension for 30 minutes to create nuclei on the Si surface. And then the Si wafer was cleaned with acetone, ethyl alcohol, and deionized water using an ultrasonic cleaner, and then dried on a hot plate. Thereafter, the as treated Si wafer was loaded into a HFCVD system for the growth of a 48 μm-thick polycrystalline CVD diamond film. During the growth process, the temperature of a Ta filament was 2200°C and it was kept at a distance of 15 mm from the substrate with 750°C. The flow ratio of hydrogen ($H_2$) and methane ($CH_4$) was 15:1 under a low working pressure of 3 KPa, which resulted in a growth rate of ~600 nm/h.

**Sample Characterization**

The purity and homogeneity of the CVD diamond are guaranteed by scanning electron microscopy images (Figs. S1(a-b) in Supplementary Note 1). According to the atomic force microscope image in Fig. S1(c), the average grain size of the CVD diamond is ~400 nm and the average grain height is ~50 nm. The free-standing CVD diamond consists of a conducting surface (~50 nm) that contains rich gain boundaries and an insulating bulk layer (~48 μm) as illustrated in Fig. 1(a). The vibrational mode and bonding configuration in the CVD diamond are characterized via Raman spectrum and X-ray photoelectron spectrum in Figs. S2(a-c), indicating that the high-uniform and impurity-free CVD diamond is successfully prepared.

**THz emission spectroscopy**

The THz emission spectroscopy with a transmission configuration is depicted in Fig. S3 (see Supplementary Note 2). A pump pulse with 800 nm central wavelength is used to illuminate the CVD diamond, from which the emitted parallel ($E_{THz-p}$) and perpendicular ($E_{THz-s}$) components of the THz electric field are obtained via a pair of wire-grid polarizers. According to the THz emission schematic in Fig. 1(a), the laboratory and crystalline coordinates of the CVD diamond are labeled by (*XYZ*) and (*X'Y'Z'*), respectively. The azimuthal angle ($\gamma$) of sample and the polarization angle ($\varphi$) of the incident light can be changed by rotating the sample holder (within *X'-Y'* plane) and a half-wave plate, respectively. The incident angle $\theta$ is tunable via rotating the sample holder within *X'-Z'* plane.

**The optical pump-infrared probe spectroscopy**

A regenerative amplifier (Spectra-physics, Spitfire) with 800 nm central wavelength is employed as the laser source. The emitted pulse is divided into two beams. One with 30% output energy is used as the pump pulse, while the other with 70% output energy is imported into an optical parametric amplifier system. In this system, a generated white light continuum has been amplified for two times, then the mid-infrared light with central wavelength in the range of 2.6~11 μm is selectively obtained. The mid-infrared light is used as the probe beam, which is overlapped with the pump beam on samples with a 100 μm spot size. More details have been introduced in the previous work[66].

**Acknowledgements**

This work was supported by National Natural Science Foundation of China under Grant Nos.12261141662, 12004310, 12074311, 61888102, 11974386, 62174179, 12074420, 11974020, U21A20140, 61905274, 22073111 and 922050307, the National Key Research and Development Program of China under Grant No. 2021YFA1400700, the Beijing Municipal Science & Technology Commission, Administrative Commission of Zhongguancun Science Park under Grant No. Z211100004821009，the Strategic Priority Research Program of Chinese Academy of Sciences (CAS) under Grant Nos. XDB33000000 and XDB28000000, the Key Research Program of Frontier Sciences of CAS under Grant Nos. QYZDJ-SSWSLH042 and XDPB22.

**Author contributions**

X.Y.X. and W.Y.D. performed the THz emission spectroscopy measurements and wrote the manuscript; Y.Y.H., B.G.Q., and X.L.X. conceived the idea and designed the research program; Z.L. and L.P.Z. performed theoretical analysis of photon drag effect; B.G.Q., G.Q.L. and C.Z.G. grew the films and the corresponding characterization; W.T., Y.X.Z., Y.L., and Y.L.L. performed the carrier dynamics measurements and corresponding theoretical analysis. all authors commented the manuscript and X.L.X. led the project.

**Competing interests**
The authors declare no competing interests.

**References**

1. Hirsch, A. The era of carbon allotropes. *Nat Mater* **9**, 868-871 (2010).

2. Coville, N. J., Mhlanga S. D., Nxumalo E. N. & Shaikjee A. A review of shaped carbon nanomaterials : review article. *S Afr J Sci* **107**, 1-15 (2011).

3. Ward, A., Broido D. A., Stewart D. A. & Deinzer G. Ab initio theory of the lattice thermal conductivity in diamond. *Phys Rev B* **80**, 125203 (2009).

4. Aharonovich, L., Greentree A. D. & Prawer S. Diamond photonics. *Nat Photonics* **5**, 397-405 (2011).

5. May, P. W. & Thompson J. M. T. Diamond thin films: a 21st-century material. *Phil Trans R Soc Lond A* **358**, 473-495 (2000).

6. Almeida, J.M. P. et al. Nonlinear optical spectrum of diamond at femtosecond regime. *Sci Rep* **7**, 14320 (2017).

7. Fisher, K. A. G. et al. Frequency and bandwidth conversion of single photons in a room-temperature diamond quantum memory. *Nat Commun* **7**, 11200 (2016).

8. Kippenberg, T. J., Holzwarth R. & Diddams S. A. Microresonator-Based Optical Frequency Combs. *Science* **332**, 555-559 (2011).

9. Chen, H. et al. Enhanced stimulated Brillouin scattering utilizing Raman conversion in diamond. *Appl Phys Lett* **120**, (2022).

10. Hausmann, B. J. M., Bulu I., Venkataraman V., Deotare P. & Lončar M. Diamond nonlinear photonics. *Nat Photonics* **8**, 369-374 (2014).

11. Laguna, V. M. F. & Panoiu N. C. Nonlinear optics in diamond-fin photonic nanowires: soliton formation and frequency comb generation. *Opt Express* **30**, 36368-36378 (2022).

12. Zouboulis, E. S. & Grimsditch M. Raman scattering in diamond up to 1900 K. *Phys Rev B* **43**, 12490-12493 (1991).

13. Abulikemu, A., Kainuma Y., An T. & Hase M. Second-Harmonic Generation in Bulk Diamond Based on Inversion Symmetry Breaking by Color Centers. *ACS Photonics* **8**, 988-993 (2021).

14. Jia, Lei, Song Yakai, Si M. S., Yao J. L. & Zhang G. P. Even-order harmonics in the nitrogen vacancy center in diamond from intertwined intraband and interband

transitions. *Phys Rev B* **105**, 214309 (2022).

15. Becker, J. N. & Becher C. Coherence Properties and Quantum Control of Silicon Vacancy Color Centers in Diamond. *Physica Status Solidi (A)* **214**, 1700586 (2017).

16. Boolakee, T. et al. Light-field control of real and virtual charge carriers. *Nature* **605**, 251-255 (2022).

17. Strother, T. et al. Photochemical Functionalization of Diamond Films. *Langmuir* **18**, 968-971 (2002).

18. Prasoon, K. S. et al. Diamond Integrated Quantum Nanophotonics: Spins, Photons and Phonons. *J Lightwave Technol* **40**, 7538-7571 (2022).

19. Ekimov, E. A. et al. Superconductivity in diamond. *Nature* **428**, 542-545 (2004).

20. Chen, X., Mohr M., Bruhne K. & Fecht H. J. Highly Conductive Nanocrystalline Diamond Films and Electronic Metallization Scheme. *Materials* **14**, 4484 (2021).

21. Sidorov, V. A. & Ekimov E. A. Superconductivity in diamond. *Diamond Relat Mater* **19**, 351-357 (2010).

22. Gajewski, W. et al. Electronic and optical properties of boron-doped nanocrystalline diamond films. *Phys Rev B* **79**, 045206 (2009).

23. Rath, P., Ummethala S., Nebel C. E. & Pernice W. H. P. Diamond as a material for monolithically integrated optical and optomechanical devices. *Physica Status Solidi (A)* **212**, 2385-2399 (2015).

24. Trojánek, F., Žídek K., Dzurňák B., Kozák M. & Malý P. Nonlinear optical properties of nanocrystalline diamond. *Opt Express* **18**, 1349-1357 (2010).

25. Kozák, M., Trojánek F. & Malý P. Large prolongation of free-exciton photoluminescence decay in diamond by two-photon excitation. *Opt Lett* **37**, 2049-2051 (2012).

26. Girolami, M. et al. Graphite distributed electrodes for diamond-based photon-enhanced thermionic emission solar cells. *Carbon* **111**, 48-53 (2017).

27. He, Y. H. et al. High-Order Shift Current Induced Terahertz Emission from Inorganic Cesium Bromine Lead Perovskite Engendered by Two-Photon

Absorption. *Adv Funct Mater* **29**, 1904694 (2019).

28. Polónyi, G. et al. High-energy terahertz pulses from semiconductors pumped beyond the three-photon absorption edge. *Opt Express* **24**, 23872-23882 (2016).

29. Zhai, D. W., Hérault E., Garet F. & Coutaz J. L. Terahertz generation from ZnTe optically pumped above and below the bandgap. *Opt Express* **29**, 17491-17498 (2021).

30. Zhu, L. et al. Enhanced polarization-sensitive terahertz emission from vertically grown graphene by a dynamical photon drag effect. *Nanoscale* **9**, 10301-10311 (2017).

31. Maysonnave, J. et al. Terahertz Generation by Dynamical Photon Drag Effect in Graphene Excited by Femtosecond Optical Pulses. *Nano Lett* **14**, 5797-5802 (2014).

32. Huang, S. et al. Terahertz emission from vertically aligned multi-wall carbon nanotubes and their composites by optical excitation. *Carbon* **132**, 335-342 (2018).

33. Huang, Y. Y. et al. Surface Optical Rectification from Layered $MoS_2$ Crystal by THz Time-Domain Surface Emission Spectroscopy. *ACS Appl Mater Interfaces* **9**, 4956-4965 (2017).

34. Cheng, L. et al. Giant photon momentum locked THz emission in a centrosymmetric Dirac semimetal. *Sci Adv* **9**, eadd7856 (2023).

35. Chen, Q. & Zhang X. C. Polarization modulation in optoelectronic generation and detection of terahertz beams. *Appl Phys Lett* **74**, 3435-3437 (1999).

36. Wu, X. J., Xu X. L., Lu X. C. & Wang L. Terahertz emission from semi-insulating GaAs with octadecanthiol-passivated surface. *Appl Surf Sci* **279**, 92-96 (2013).

37. Zhang, L. H. et al. Polarized THz Emission from In-Plane Dipoles in Monolayer Tungsten Disulfide by Linear and Circular Optical Rectification. *Adv Opt Mater* **7**, 1801314 (2019).

38. Fan, Z. Y. et al. Terahertz Surface Emission from $MoSe_2$ at the Monolayer Limit.

*ACS Appl Mater Interfaces* **12**, 48161-48169 (2020).

39. Najar, H. et al. High quality factor nanocrystalline diamond micromechanical resonators limited by thermoelastic damping. *Appl Phys Lett* **104**, 151903 (2014).

40. Chang, J. W. et al. Coherent Elliptically Polarized Terahertz Wave Generation in WSe$_2$ by Linearly Polarized Femtosecond Laser Excitation. *J Phys Chem Lett* **12**, 10068-10078 (2021).

41. Osterhoudt, G. B. et al. Colossal mid-infrared bulk photovoltaic effect in a type-I Weyl semimetal. *Nat Mater* **18**, 471-475 (2019).

42. Obraztsov, P. A. et al. Photon-drag-induced terahertz emission from graphene. *Phys Rev B* **90**, 241416 (2014).

43. Apretna, T. et al. Coherent THz wave emission from HgTe quantum dots. *Appl Phys Lett* **121**, 251101 (2022).

44. Cao, X. Q. et al. Interplay between Ultrafast Shift Current and Ultrafast Photon Drag Current in Tellurium Nanotubes. *ACS Photonics* **9**, 3144-3155 (2022).

45. Bahk, Y. M. et al. Plasmon enhanced terahertz emission from single layer graphene. *ACS Nano* **8**, 9089-9096 (2014).

46. Zhu, L. P. et al. Circular-Photon-Drag-Effect-Induced Elliptically Polarized Terahertz Emission from Vertically Grown Graphene. *Phys Rev Appl* **12**, 044063 (2019).

47. Sun, Ying‑Chieh, Gai Huadong & Voth Gregory A. Vibrational energy relaxation dynamics of C–H stretching modes on the hydrogen‑terminated H/C(111)1×1 surface. *The Journal of Chemical Physics* **100**, 3247-3251 (1994).

48. Cheng, C. L., Chen C. F., Shaio W. C., Tsai D. S. & Chen K. H. The CH stretching features on diamonds of different origins. *Diamond Relat Mater* **14**, 1455-1462 (2005).

49. Howell, D. et al. Automated FTIR mapping of boron distribution in diamond. *Diamond Relat Mater* **96**, 207-215 (2019).

50. Wang, L. J. et al. Ellipsometric analysis and optical absorption characterization of nano-crystalline diamond film. *Trans Nonferrous Met Soc China* **16**, s289-s292

(2006).

51. Gao, X. et al. Integrated wafer-scale ultra-flat graphene by gradient surface energy modulation. *Nat Commun* **13**, 5410 (2022).

52. Jan, L. et al. High Carrier Mobility in Single-Crystal Plasma-Deposited Diamond. *Science* **297**, 1670-1672 (2002).

53. Wang, K. et al. High-Contrast Polymorphic Luminogen Formed through Effect of Tiny Differences in Intermolecular Interactions on the Intramolecular Charge Transfer Process. *Adv Opt Mater* **8**, 2000436 (2020).

54. Sun, Y. C., Gai H. D. & Voth G. A. Vibrational energy relaxation dynamics of C–H stretching modes on the hydrogen-terminated H/C(111)1×1 surface. *J Chem Phys* **100**, 3247-3251 (1994).

55. Jahan, M. D., Shivaraman S., Chandrashekhar M., Michael G. S. & Rana F. Measurement of Ultrafast Carrier Dynamics in Epitaxial Graphene. *Appl Phys Lett* **1081**, 604 (2008).

56. Achatz, P. et al. Optical properties of nanocrystalline diamond thin films. *Appl Phys Lett* **88**, 101908 (2006).

57. Nebel, C. E. Electronic properties of CVD diamond. *Semicond Sci Technol* **18**, S1 (2003).

58. Choi, H. et al. Broadband electromagnetic response and ultrafast dynamics of few-layer epitaxial graphene. *Appl Phys Lett* **94**, 172102 (2009).

59. Dawlaty, J. M., Shivaraman S., Chandrashekhar M., Rana F. & Spencer M. G. Measurement of ultrafast carrier dynamics in epitaxial graphene. *Appl Phys Lett* **92**, 042116 (2008).

60. Huang, L. B. et al. Ultrafast Transient Absorption Microscopy Studies of Carrier Dynamics in Epitaxial Graphene. *Nano Lett* **10**, 1308-1313 (2010).

61. Newson, R. W., Dean J., Schmidt B. & Driel H. M. Ultrafast carrier kinetics in exfoliated graphene and thin graphite films. *Opt Express* **17**, 2326-2333 (2009).

62. Sun, D. et al. Ultrafast Relaxation of Excited Dirac Fermions in Epitaxial Graphene Using Optical Differential Transmission Spectroscopy. *Phys Rev Lett*

**101**, 157402 (2008).

63. Li, D. et al. Ultrafast Dynamics of Defect-Assisted Auger Process in PdSe2 Films: Synergistic Interaction between Defect Trapping and Auger Effect. *J Phys Chem Lett* **13**, 2757-2764 (2022).

64. Suo, P. et al. Terahertz Emission on the Surface of a van der Waals Magnet CrSiTe$_3$. *Laser Photonics Rev* **14**, 2000025 (2020).

65. Obraztsov, P. A. et al. All-optical control of ultrafast photocurrents in unbiased graphene. *Sci Rep* **4**, 4007 (2014).

66. Zhang, J. et al. Identifying and Modulating Accidental Fermi Resonance: 2D IR and DFT Study of 4-Azido-l-phenylalanine. *J Phys Chem B* **122**, 8122-8133 (2018).